\documentclass[12pt,a4paper]{article}

\usepackage[dvips]{graphicx}
\usepackage{cite,float}
\usepackage{latexsym,amssymb,amsmath}
\usepackage{url,calc,times}

\newcommand{\shc}{spherical harmonics coefficients}
\newcommand{\xavr}[1]{<\!\! #1\!\!>}
\newcommand{\xtavr}[1]{<\!\!<\!\! #1\!\!>\!\!>}
\newcommand{\comment}[1]{}
\newcommand{\bequ}{\begin{equation}}
\newcommand{\eequ}[1]{\label{#1}\end{equation}}
\newcommand\eq[1] {(\ref{#1})}
\newcommand{\bfm}[1]{\boldsymbol{#1}}
\DeclareMathOperator*{\floor}{floor}

\title{The Rate of Expansion of Spherical Flames} 
\author{V. Karlin\thanks{University of Central Lancashire, 
Preston, UK}\hspace{3mm} and\hspace{3mm} 
G. Sivashinsky\thanks{Tel Aviv University, Tel Aviv, Israel}}
\date{}

\begin{document}
\maketitle
\thispagestyle{empty} 

\begin{abstract}
In this paper we investigate the acceleration of the expansion of 
premixed spherical flames and evolution of the cellular 
patterns on their surfaces. An asymptotic model is used for 
the simulations and a spectral numerical algorithm is 
employed to study flames over large time intervals. 
Numerous numerical experiments indicate that for large 
enough time intervals the acceleration of two-dimensional 
expanding flame ceases and the expansion rate stabilizes 
to a value significantly exceeding the burning rate. The 
importance of the effect of forcing was also confirmed and the 
validity of sectorial simulations of closed flames was 
studied in order to justify prospective use of the Fourier 
spectral model for the three-dimensional spherical flames. 
\end{abstract}

\section{Introduction} 
In \cite{Gostintsev-Istratov-Shulenin88} extensive experimental 
data on premixed expanding spherical flames have been analysed 
and it was concluded that starting from a certain moment the 
averaged flame radius $\xavr{r}$ grows as $\xavr{r}\ \propto t^{3/2}$ 
rather than $\xavr{r}\ \propto t$. The phenomenon was linked to the 
cellularization of flames, which was well known from experiments 
too\comment{, see e.g. \cite{Groff82}}. Indeed, the appearance of 
cellular patterns increases the flame surface area, hence the 
fuel consumption, and hence the averaged flame expansion rate.

The cellularization of flame fronts was, in its turn, associated 
with intrinsic combustion instabilities. The effect of the hydrodynamic 
combustion instability on expanding spherical flames was studied in 
\cite{Istratov-Librovich69} using the linear perturbation theory 
combined with phenomenological assumptions. Later, the approach 
was further improved and freed from the phenomenological assumptions 
\cite{Bechtold-Matalon87}. These linearized solutions confirmed the 
onset of the instability of the flame front but could not quantify 
its cellularization and acceleration because the latter phenomena 
are essentially nonlinear. 

A simple, yet physically reasonable, nonlinear model of
hydrodynamically unstable planar flames was suggested in
\cite{Sivashinsky77a}. In \cite{Filyand-Sivashinsky-Frankel94} it 
was extended to expanding spherical flames and studied numerically 
confirming that there is a time instance $t_{*}$, such that 
the flame expansion rate behaves like $\xavr{r}\ \propto t^{3/2}$ 
for $t>t_{*}$. The same model was also suggested in \cite{Joulin94a}, 
where similar to the planar flames \cite{Thual-Frisch-Henon85}, 
analytical pole solutions were obtained and studied as well.

Because of their physical origins simple nonlinear models of 
expanding flames \cite{Filyand-Sivashinsky-Frankel94,Joulin94a} 
are expected to be valid only locally. Apparently, the results 
obtained when applying them 
to the whole flame are instructive indeed, but still inconclusive 
and cannot be accepted as the adequate theoretical model of 
cellularization and acceleration. A physically consistent global 
model of flames of any smooth enough geometry was developed 
in \cite{Frankel90}. Mathematically, the approach projects the 
governing equations to the flame surface reducing the mathematical 
dimension of the problem by one. However, the resulting equation 
is still extremely costly from the computational point of view 
and only two-dimensional simulations have been carried out so far. 

A compromise between universality and computability was suggested in 
\cite{DAngelo-Joulin-Boury00}, where consideration was limited to a 
narrow but still very practical case of flames which do not deviate 
from the spherical ones significantly. On the technical side the model 
combines the operator of the linearized problem obtained in 
\cite{Istratov-Librovich69} for the expanding spherical flame in 
terms of spherical harmonics expansions and a Huygens type 
nonlinearity specific to the local nonlinear model 
\cite{Filyand-Sivashinsky-Frankel94,Joulin94a}. 
Physically, model \cite{DAngelo-Joulin-Boury00} is consistent with 
\cite{Frankel90} and is robust and plausible enough to simulate the 
cellularization of expanding spherical flames in three spatial 
dimensions. At the time of writing of this paper, the flame sizes 
we were able to reach in our computations do not significantly exceed those 
reported in \cite{DAngelo-Joulin-Boury00} and are 
not large enough to match our two-dimensional calculations. However, 
our investigations show that numerical studies of the expanding 
three-dimensional flames on the time scales required for comparison 
with the two-dimensional calculations are possible.

In the following sections we specify the mathematical models 
and numerical algorithms to solve them. In Section \ref{Results} 
we report our results on flame front behaviour on long time 
intervals and on the effect of external forcing. Also, we 
present our attempts to simulate the three-dimensional flames 
and assess the possibility of simulation of closed flames 
via their finite segments.

\section{Mathematical Models}

Let us consider an expanding flame front and assume that its surface
is close enough to a sphere and that every point on the flame surface
is uniquely defined by its distance $r=r(\theta,\phi,t)$
from the origin for $0\le\theta\le\pi$, $0\le\phi\le 2\pi$, 
and $t>0$. It is convenient to represent such a flame as a 
perturbation $\Phi(\theta,\phi,t)$ of a spherical surface of a 
reference radius $r_{0}(t)$, i.e.
$r(\theta,\phi,t)=r_{0}(t)+\Phi(\theta,\phi,t)$.
Then, the Fourier image of the governing equation of the flame 
front evolution in the nondimensional notations suggested in 
\cite{Filyand-Sivashinsky-Frankel94,Joulin94a} can be written 
as
\[
\frac{d\widetilde{\Phi}_{k}}{dt}
=\left\{-\frac{\theta_{\pi}^{2}}{[r_{0}(t)]^{2}}|k|^{2}
+\frac{\gamma\theta_{\pi}}{2r_{0}(t)}|k|\right\}\widetilde{\Phi}_{k}
\]
\bequ
-\frac{\theta_{\pi}^{2}}{2[r_{0}(t)]^{2}}\sum\limits_{l=-\infty}^{\infty}
l(k-l)\widetilde{\Phi}_{l}\widetilde{\Phi}_{k-l}
+\widetilde{f}_{k}(t).
\eequ{SivCurv1b}
Here $|k|<\infty$, $t>0$, $\widetilde{f}_{k}(t)$ are the Fourier 
components of the properly scaled upstream perturbations of the 
unburnt gas velocity field $f(\phi,t)$, and initial values of 
$\widetilde{\Phi}_{k}(0)=\widetilde{\Phi}_{k}^{(0)}$ are given. 
By construction, equation \eq{SivCurv1b} holds in the sector 
$0\le\phi\le 2\pi/\theta_{\pi}$ with a large enough integer 
$\theta_{\pi}$.

Models of \cite{Filyand-Sivashinsky-Frankel94} and \cite{Joulin94a} 
differ by an additive term proportional to $\delta_{0,k}/r_{0}(t)$. 
This term adds just $O(\ln t)$ to $\widetilde{\Phi}_{k}$ for 
$k=0$ only. This is not essential and the term is not included in 
our conception. 

Equation \eq{SivCurv1b} was obtained as a local model of a 
curved expanding flame. However, we will use it globally, on 
the whole flame surface, with $\theta_{\pi}=1$. In order to 
justify such an action, let us first note that if transformed 
back to the physical space, the equation takes the form
\[
\frac{\partial\Phi(\phi,t)}{\partial t}
=\frac{1}{[r_{0}(t)]^{2}}\frac{\partial^{2}\Phi(\phi,t)}{\partial\phi^{2}}
+\frac{\gamma\theta_{\pi}}{2\pi r_{0}(t)}\frac{\partial}{\partial\phi}
\int\limits_{0}^{2\pi/\theta_{\pi}}
\Phi(\theta,t)\cot\frac{\phi-\theta}{2}d\theta
\]
\bequ
+\frac{1}{2[r_{0}(t)]^{2}}\left[\frac{\partial\Phi(\phi,t)}
{\partial\phi}\right]^{2}+f(\phi,t),
\eequ{SivCurv1c}
which is rotation-invariant for $f(\phi,t)\equiv 0$ and 
$\theta_{\pi}=1$.

On the other hand, we may rewrite the geometrically invariant 
equation obtained in \cite{Frankel90} in the coordinate form as 
follows:
\[
\frac{\partial r}{\partial t}
=\sqrt{1+\left(\frac{r_{\phi}}{r}\right)^{2}}\Bigg\{1
-\epsilon\frac{r^{2}+2(r_{\phi})^{2}-rr_{\phi\phi}}
{[r^{2}+(r_{\phi})^{2}]^{3/2}}
\]
\bequ
\left.-\frac{\gamma}{2}
\left[1+\frac{1}{\pi}
\int\limits_{0}^{2\pi}
p(\phi,\theta)d\theta\right]\right\}, 
\eequ{int5}
where
\bequ
p(\phi,\theta)=\left\{
\begin{array}{ll}
\dfrac{[\bfm{r}(\phi)-\bfm{r}(\theta)]\cdot\bfm{n}(\phi)}
{|\bfm{r}(\phi)-\bfm{r}(\theta)|^{2}}
\sqrt{[r(\theta)]^{2}+(r_{\theta})^{2}} & |\theta-\phi|>0,\\
 & \\
-\dfrac{[r(\phi)]^{2}+2(r_{\phi})^{2}
-r(\phi)r_{\phi\phi}}{2\{[r(\phi)]^{2}+(r_{\phi})^{2}\}} 
 & |\theta-\phi|=0,\\
\end{array}\right.
\eequ{int5a}
\bequ
\bfm{n}(\phi)=\frac{[-(r_{\phi}\sin\phi+r\cos\phi),(r_{\phi}\cos\phi-r\sin\phi)]}
{\sqrt{r^{2}+(r_{\phi})^{2}}},
\eequ{int5b}
and
\bequ
\bfm{r}(\phi)=[r(\phi)\cos\phi,r(\phi)\sin\phi].
\eequ{int5c}
Assuming again that $r(\phi,t)=r_{0}(t)+\Phi(\phi,t)$ and 
linearizing all but the Huygens terms, one arrives to the 
equation 
\[
\frac{\partial\Phi(\phi,t)}{\partial t}
=\frac{\epsilon}{[r_{0}(t)]^{2}}\frac{\partial^{2}\Phi(\phi,t)}{\partial\phi^{2}}
+\frac{\gamma}{2\pi r_{0}(t)}\frac{\partial}{\partial\phi}\int\limits_{0}^{2\pi}
\Phi(\theta,t)\cot\frac{\phi-\theta}{2}d\theta
\]
\[
+\frac{1}{2[r_{0}(t)]^{2}}\left[\frac{\partial\Phi(\phi,t)}
{\partial\phi}\right]^{2}
+1-\frac{dr_{0}(t)}{dt}-\frac{\epsilon}{r_{0}(t)}
\]
\bequ
+\left\{\frac{\epsilon}{[r_{0}(t)]^{2}}-\frac{\gamma}{r_{0}(t)}\right\}\Phi(\phi,t)
+\frac{\gamma}{2\pi r_{0}(t)}\int\limits_{0}^{2\pi}\Phi(\theta)d\theta. 
\eequ{siv1}
which differs from \eq{SivCurv1c}, for $\theta_{\pi}=1$, by a 
few non-essential terms only. 

Comparison of \eq{SivCurv1c} and \eq{siv1} reveals the detailed 
relationship between the simplified model 
\cite{Filyand-Sivashinsky-Frankel94,Joulin94a} and the 
comprehensive one \cite{Frankel90}. In particular, it validates 
use of \eq{SivCurv1c} as a global model of the whole spherical 
flame as long as the perturbations remain small enough. 

Equations of the three-dimensional model \cite{DAngelo-Joulin-Boury00} 
can be written in terms of the spherical harmonics expansion 
coefficients 
\[
\widetilde{\Phi}_{n,m}(t)=\int\limits_{0}^{2\pi}
\int\limits_{0}^{\pi}
\Phi(\theta,\phi,t)\overline{Y_{n,m}(\theta,\phi)}
\sin\theta d\theta d\phi,\qquad |n|<\infty,\ |m|\le|n|
\] 
of $\Phi(\theta,\phi,t)$ as
\bequ
\frac{d\widetilde{\Phi}_{n,m}(t)}{dt}=\omega(n,t)
\widetilde{\Phi}_{n,m}(t)
+\frac{1}{2[r_{0}(t)]^{2}}\widetilde{N}_{n,m}(t)
+\widetilde{f}_{n,m}(t).
\eequ{GovEq3}
Here $|n|,|m|<\infty$, $t>0$, $\widetilde{f}_{n,m}(t)$ are the 
\shc\ of the properly scaled upstream perturbations 
of the unburnt gas velocity field, and initial values of 
$\widetilde{\Phi}_{n,m}(0)=\widetilde{\Phi}_{n,m}^{(0)}$ are given. 
The expression for the linear response
\bequ
\omega(n,t)=\frac{2n(n-1)}{(2n+1)r_{0}(t)}
-\frac{(n-1)(n+2)}{r_{0}^{2}(t)},\quad
\eequ{GovEq3b}
emerges from the analysis of \cite{Istratov-Librovich69} and 
$\widetilde{N}_{n,m}(t)$ are the \shc\ of the nonlinear Huygens 
term 
\bequ
\mathcal{N}(\Phi)=
\|\nabla_{S}\Phi\|^{2}=\left(\frac{\partial\Phi}{\partial\theta}\right)^{2}
+\frac{1}{(\sin\theta)^{2}}\left(\frac{\partial\Phi}{\partial\phi}\right)^{2}.
\eequ{GovEq2b}

Choice of $r_{0}(t)$ in the models introduced in this Section 
may be based on a variety of principles. In equation \eq{siv1}, 
which governs small perturbations of \eq{int5}-\eq{int5c}, choice 
of $r_{0}(t)$ is more or less arbitrary. However
\bequ
r_{0}(t)=\frac{1}{2\pi}\int\limits_{0}^{2\pi}r(\phi,t)d\phi
\eequ{r0a}
is probably the most appropriate one because it minimizes the 
perturbations $r(t)-r_{0}(t)=\Phi(\phi,t)$. On the other hand, 
equation 
\eq{SivCurv1b}, or \eq{SivCurv1c}, was obtained as that one, 
which governs perturbations of a uniform and steadily 
expanding spherical flame $r(t)\equiv t$. Therefore, choice of 
\bequ
r_{0}(t)\equiv t
\eequ{r0b}
is more reasonable in this case, as this is the solution around 
which the linearization is performed. Equation \eq{SivCurv1b} 
was already considered with \eq{r0a} in 
\cite{Filyand-Sivashinsky-Frankel94}. Here we complement the 
analysis of \eq{SivCurv1b} by considering it with \eq{r0b}.

\section{Computational Algorithms}

System \eq{SivCurv1b} is solved numerically by neglecting 
the harmonics of orders higher than a finite integer number 
$K>0$. Then, the nonlinearity can be represented as a circular 
convolution and evaluated effectively with the FFT. Also, we 
found that the stability of the numerical integration scheme 
suggested in \cite{Filyand-Sivashinsky-Frankel94} can be improved 
significantly by calculating the contribution from the linear terms 
in \eq{SivCurv1b} analytically. Thus, the linear terms, i.e. the 
source of physical instability, are tackled exactly and only the 
nonlinear ones, with the dumping effect, are approximated. 
This improvement allowed us to continue the calculations for up 
to ten times further in time than in 
\cite{Filyand-Sivashinsky-Frankel94}.

Using the notation 
\bequ
g_{k}[\widetilde{\Phi}]=-\frac{\theta_{\pi}^{2}}{2}
\sum\limits_{l=-K}^{K}
l(k-l)\widetilde{\Phi}_{l}\widetilde{\Phi}_{k-l}
-V\delta_{k,0}+f_{k}(t),
\eequ{SivCurv10b}
equation \eq{SivCurv1b} can be written as
\bequ
\frac{d\widetilde{\Phi}_{k}}{dt}
=\left(-\frac{\theta_{\pi}^{2}}{r_{0}^{2}(t)}|k|^{2}
+\frac{\gamma\theta_{\pi}}{2r_{0}(t)}|k|\right)
\widetilde{\Phi}_{k}
+\frac{1}{r_{0}^{2}(t)}g_{k},\qquad |k|\le K.
\eequ{SivCurv10c}
and we will search for its solutions in the form
\bequ
\widetilde{\Phi}_{k}(t)=C_{k}(t)Y_{k}(t),
\eequ{SivCurv11a}
where $Y(t)$ is the solution of the uniform equation.

Straightforward evaluations yield 
\bequ
Y_{k}(t)=e^{\Delta\omega_{k}(t)},\qquad 
\Delta\omega_{k}(t)=-\theta_{\pi}^{2}|k|^{2}
\int\limits_{t_{0}}^{t}\frac{dt}{r_{0}^{2}(t)}
+\frac{1}{2}\gamma\theta_{\pi}|k|
\int\limits_{t_{0}}^{t}\frac{dt}{r_{0}(t)}.
\eequ{SivCurv11b}
Then, the equation for $C_{k}(t)$ is
\bequ
\frac{dC_{k}}{dt}
=r_{0}^{-2}(t)e^{\theta_{\pi}^{2}|k|^{2}
\int\limits_{t_{0}}^{t}\frac{dt}{r_{0}^{2}(t)}
-\frac{1}{2}\gamma\theta_{\pi}|k|
\int\limits_{t_{0}}^{t}\frac{dt}{r_{0}(t)}}g_{k},
\eequ{SivCurv11d}
where $t_{0}$ is an arbitrary real of which the final result 
\eq{SivCurv11a} does not depend.

In order to integrate the ODE's \eq{SivCurv11d} over 
the interval $[t_{n},t_{n}+\Delta t]$ or 
$[t_{n}-\Delta t,t_{n}+\Delta t]$, we interpolate values 
of $g_{k}(t)\equiv g_{k}[\widetilde{\Phi}]$ polynomially: 
\bequ
g_{k}(t)=\sum\limits_{\mu=0}^{m_{p}}G_{k,n,\mu}t^{\mu},\qquad 
t\in[t_{n}-\Delta t,t_{n}+\Delta t].
\eequ{poly0}
For example, the first order extrapolation gives
\bequ
g_{k}(t)\equiv G_{k,n,0}=g_{k}(t_{n}),
\eequ{poly1}
and the second one results in
\bequ
g_{k}(t)=g_{k}(t_{n})\frac{t-t_{n}+\Delta t}{\Delta t}
-g_{k}(t_{n-1})\frac{t-t_{n}}{\Delta t}
=G_{k,n,0}+G_{k,n,1}t,
\eequ{poly2}
where
\bequ
G_{k,n,0}=\frac{-g_{k}(t_{n})(t_{n}-\Delta t)
+g_{k}(t_{n-1})t_{n}}{\Delta t},\quad
G_{k,n,1}=\frac{g_{k}(t_{n})-g_{k}(t_{n-1})}{\Delta t}.
\eequ{poly4}
In what follows we provide formulas for the first order extrapolation 
only. Formulas of higher orders of accuracy are a bit bulky, but can 
be obtained straightforwardly.

The interpolation results, in general, in 
\bequ
\widetilde{\Phi}_{k}(t_{n}+\Delta t)
=e^{\Delta\omega_{k}(t_{n}+\Delta t)}
\left[\widetilde{\Phi}_{k}(t_{n})+\sum\limits_{\mu=0}^{m_{p}}
\varkappa_{k,n,\mu}G_{k,n,\mu}\right]
\eequ{SivCurv11i}
with
\bequ
\Delta\omega_{k}(t)
=-\theta_{\pi}^{2}|k|^{2}
\int\limits_{t_{n}}^{t}\frac{dt}{r_{0}^{2}(t)}
+\frac{1}{2}\gamma\theta_{\pi}|k|
\int\limits_{t_{n}}^{t}\frac{dt}{r_{0}(t)},
\eequ{SivCurv11j}
and 
\bequ
\varkappa_{k,n,\mu}=\int\limits_{t_{n}}^{t_{n}+\Delta t}
\xi^{\mu}r_{0}^{-2}(\xi)e^{-\Delta\omega_{k}(\xi)}d\xi.
\eequ{SivCurv11k}

In order to move any further, we will now assume that
\bequ
r_{0}(t)\propto t^{\alpha}.
\eequ{SivCurv11l}
Then \eq{SivCurv11j} can be written as
$\Delta\omega_{k}(t)=\omega_{k}(t)-\omega_{k}(t_{n})$,
where
\bequ
\omega_{k}(t)
=\frac{\theta_{\pi}^{2}|k|^{2}t_{n}^{2}}
{r_{0}^{2}(t_{n})t}
+\frac{\gamma\theta_{\pi}|k|t_{n}\ln t}
{2r_{0}(t_{n})},\qquad \alpha=1.
\eequ{SivCurv11o}
and
\bequ
\omega_{k}(t)
=\frac{\theta_{\pi}^{2}|k|^{2}t_{n}^{2\alpha}}
{(2\alpha-1)r_{0}^{2}(t_{n})t^{2\alpha-1}}
-\frac{\gamma\theta_{\pi}|k|t_{n}^{\alpha}}
{2(\alpha-1)r_{0}(t_{n})t^{\alpha-1}},\qquad \alpha\ne 1.
\eequ{SivCurv11n}
Further, \eq{SivCurv11k} for the first order of 
approximation will become
\bequ
\varkappa_{k,n,\mu}=\frac{t_{n}^{\mu}}
{r_{0}^{2}(t_{n})}\Delta t.
\eequ{SivCurv11r}
Then, the integration formula \eq{SivCurv11i} is 
transformed into
\bequ
\widetilde{\Phi}_{k}(t_{n}+\Delta t)
=e^{\omega_{k}(t_{n}+\Delta t)-\omega_{k}(t_{n})}
\left[\widetilde{\Phi}_{k}(t_{0})+
\frac{\Delta t}{r_{0}^{2}(t_{0})}g_{k}(t_{n})\right],
\eequ{SivCurv11u}
with $\omega_{k}(t)$ given by \eq{SivCurv11o} and 
\eq{SivCurv11n}.

A computational algorithm based on numerical spherical 
harmonics transformation (SHT) \cite{Spherepack}, similar to 
\cite{DAngelo-Joulin-Boury00}, was used in this work in order to 
solve \eq{GovEq3}. In addition, the stability of the numerical 
integration scheme was improved by evaluating the contribution 
from the linear terms analytically, as shown above, 
and the code was parallelized in order to speed up the 
computations and to use larger data sets.

\section{Computational results}\label{Results}

\subsection{Saturation on the long time intervals}

\begin{figure}[ht]
\begin{centering}\hfill
\begin{tabular}{p{55mm}p{10mm}p{55mm}}
\includegraphics[height=55mm,width=55mm]{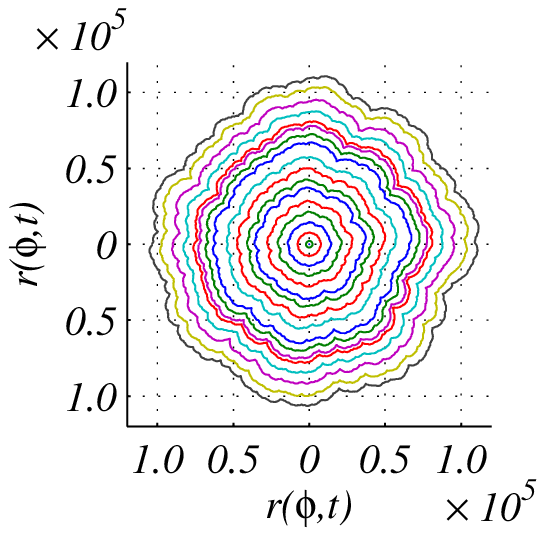}
& \phantom{a} &
\includegraphics[height=55mm,width=55mm]{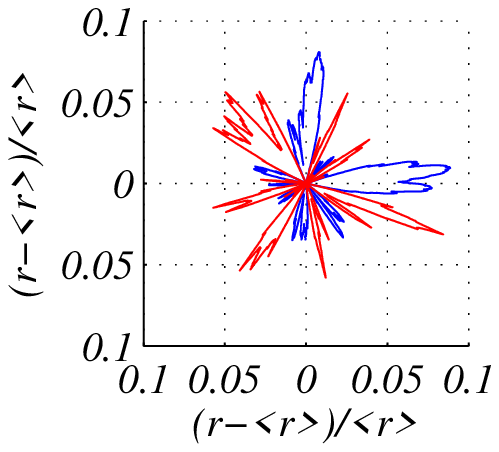}\\ 
\end{tabular}\hfill
\end{centering}
\caption{Evolution of a spherical flame governed by \eq{SivCurv1b}. 
Here values of $r(\phi,t)$ are on the left and
$[r(\phi,t)-\xavr{r}]/\!\!\xavr{r}$ for $t=7.65\times 10^{4}$ are 
on the right. Positive values of the latter are in blue and 
negative ones are in red; $\gamma=0.8$, $f(\phi,t)\equiv 0$.}
\label{sphere2d}
\end{figure}

Typical shapes of the flame fronts governed by \eq{SivCurv1b} 
over large time intervals are illustrated in Fig. 
\ref{sphere2d}. The graph of 
$[r(\phi,t)-\!\!\xavr{r}]/\!\!\!\xavr{r}$ for 
$t=7.65\times 10^{4}$ shows that the wrinkle amplitudes are 
up to 10\% of the averaged flame radius. The explicit 
forcing was not applied in this example.

\begin{figure}[ht]
\begin{centering}\hfill
\begin{tabular}{p{60mm}p{1mm}p{60mm}}
\includegraphics[height=35mm,width=60mm]{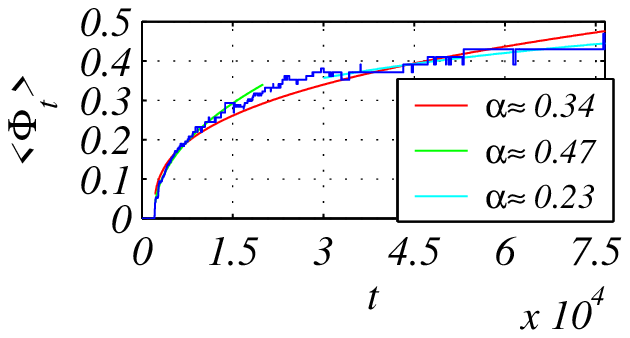} 
& \phantom{a} &
\includegraphics[height=35mm,width=60mm]{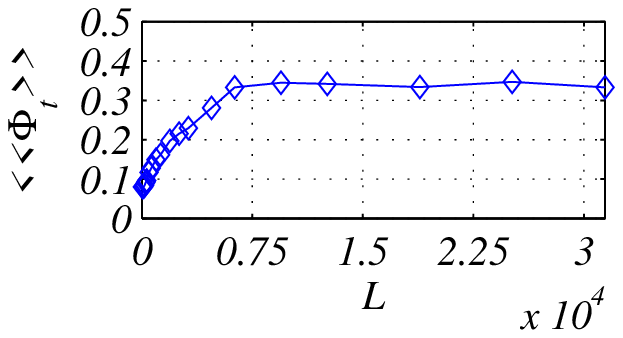}\\
\end{tabular}\hfill
\end{centering}
\caption{Averaged flame front velocity for spherical (left) and 
planar (right) flames versus time $t$ and flame size $L$ 
respectively. Here $\gamma=0.8$ and $f(\phi,t)\equiv 0$. 
Markers on the right graph show the calculated cases.}
\label{Speed}
\end{figure}

Permanent growth of the size of a spherical flame as it expands 
prompts studies of the effect of the size of a planar 
flame on its propagation speed as the first step towards the
understanding of the acceleration mechanism of the expanding 
flames. The investigation of the dynamics of planar flames shows 
that the spatially averaged flame speed 
\bequ
\xavr{\Phi_{t}}=\frac{1}{L}\int\limits_{0}^{L}
\frac{\partial\Phi_{t}}{\partial t}dx
\eequ{plsp1}
of a flame size $L$ stops growing and begins to oscillate 
irregularly around its time average $\xtavr{\Phi_{t}}$ 
for large enough $t$. A definite correlation between the size 
of the flame and its stabilized spatially averaged propagation 
speed $\xtavr{\Phi_{t}}$ was established, see e.g. 
\cite{Karlin02a}. The effect was explained by proving the high sensitivity 
of planar cellular flames to particular types of linear 
perturbations, see \cite{Joulin89b,Karlin02a,Karlin02b,Karlin04a}. 
By continuing calculations reported in \cite{Karlin02a} for even 
larger planar flames, we established that their propagation speed 
no longer grows after a certain critical flame size is 
reached, see graph on the right of Fig. \ref{Speed}. 
In this paper we are interested in extending these 
findings for planar propagating flames to the expanding ones. 
In particular, we are studying the possibility of a stabilization 
of the expansion rate for large enough time intervals, when 
the flame size grows sufficiently large.

Stabilized velocities $\xtavr{\Phi_{t}}$ of the planar flames 
and averaged velocities $\xavr{\Phi_{t}}=\frac{1}{2\pi}\int_{0}^{2\pi}
\frac{\partial\Phi_{t}}{\partial t}d\phi$ of the spherical flames 
are compared in 
Fig. \ref{Speed}. Power law approximations $(t-t_{*})^{\alpha}$ 
for the expansion rate of the spherical flame are also depicted 
there. Sudden increase of $\xavr{\Phi_{t}}$ begins from 
$t_{c}\approx 2.2\times 10^{3}$. For the whole considered time 
interval $[2.2\times 10^{3},7.65\times 10^{4}]$ the optimal 
$\alpha\approx 0.34$. For earlier times 
$t\in[2.2\times 10^{3},2.0\times 10^{4}]$, the best approximation 
is with $\alpha\approx 0.47$, i.e. almost $1/2$ as obtained in 
experiments. However, as time 
goes by, the expansion rate slows down and for 
$t\in[3.0\times 10^{4},7.65\times 10^{4}]$ we got 
$\alpha\approx 0.23$.

\begin{figure}[ht]
\begin{centering}\hfill
\begin{tabular}{p{60mm}p{1mm}p{60mm}}
\includegraphics[height=35mm,width=60mm]{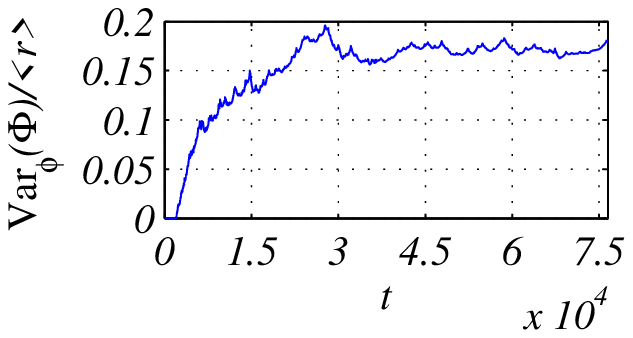} 
& \phantom{a} &
\includegraphics[height=35mm,width=60mm]{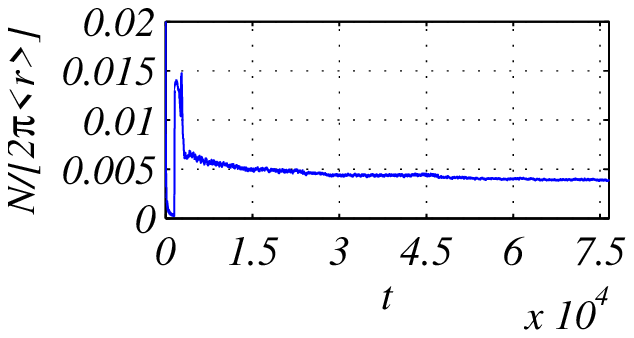}\\
\end{tabular}\hfill
\end{centering}
\caption{Ratio of the maximal amplitude of perturbation and of the 
averaged flame radius (left) and number of cells per unit length 
of the averaged flame contour (right). Here $\gamma=0.8$ and 
$f(\phi,t)\equiv 0$.}
\label{Saturation}
\end{figure}

A tendency towards stabilization of $\xavr{\Phi_{t}}$ to a 
constant for $t\rightarrow\infty$ is evident in the graph 
too. A change in morphology of the flame front at 
$t_{s}\approx 2.5\times 10^{4}$ is even more obvious in the 
graph of the variation of the perturbation of the 
averaged circular flame relatively to the radius of this 
averaged flame, i.e. of 
$\max\limits_{0\le\phi\le 2\pi}[r(\phi,t)-\!\!\xavr{r}]/\!\!\xavr{r}$, 
which is shown on the left of Fig. \ref{Saturation}. 

\begin{figure}[ht]
\begin{centering}\hfill
\begin{tabular}{p{60mm}p{1mm}p{60mm}}
\includegraphics[height=35mm,width=60mm]{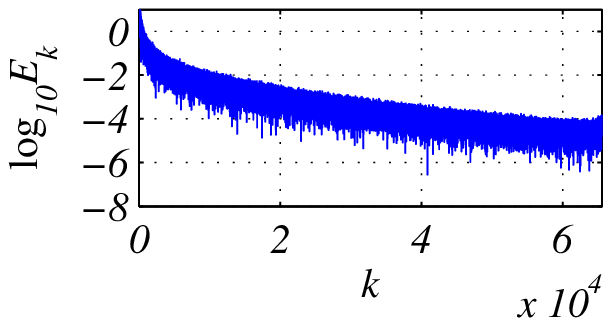} 
& \phantom{a} &
\includegraphics[height=35mm,width=60mm]{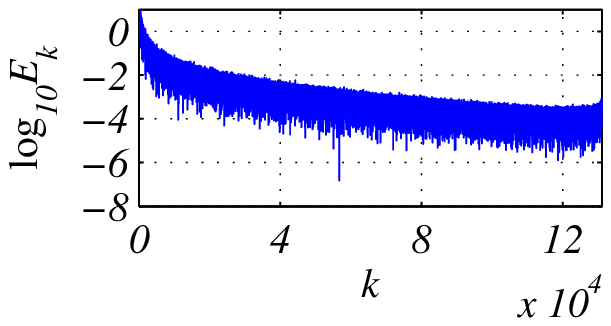}\\
\end{tabular}\hfill
\end{centering}
\caption{Spectral distributions of energy of the perturbation 
$\Phi(\phi,t)$ for $t=2.4928\times 10^{4}$ (left) and 
$t=7.65\times 10^{4}$ (right).}
\label{SpecDens}
\end{figure}

The number of cells per unit length of the averaged flame contour 
is depicted on the right of Fig. \ref{Saturation}. It stabilizes 
to a cell size of about $200$, which is exactly the same as for large 
enough planar flames. Eventually, Fig. \ref{SpecDens} illustrates 
the spectral distribution of energy of the perturbation $\Phi(\phi,t)$, 
which stabilizes by $t_{s}\approx 2.5\times 10^{4}$ too. 

\begin{figure}[ht]
\begin{centering}\hfill
\begin{tabular}{p{60mm}p{1mm}p{60mm}}
\includegraphics[height=35mm,width=60mm]{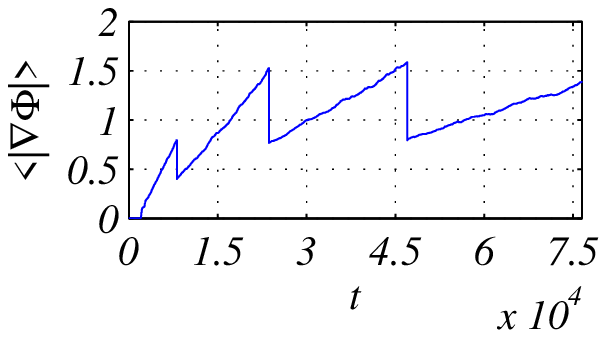} 
& \phantom{a} &
\includegraphics[height=35mm,width=60mm]{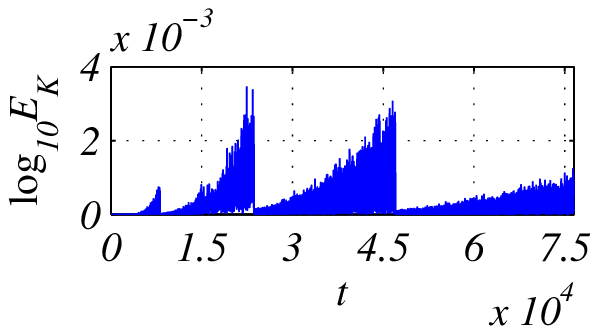}\\
\end{tabular}\hfill
\end{centering}
\caption{Averaged gradient of the perturbation (left) and energy 
of the highest order harmonic involved in the simulation (right).}
\label{BlowUp}
\end{figure}

Stabilization of the averaged perturbation gradient 
$\nabla\Phi=\partial\Phi/\partial\phi$ can also be seen on the left 
of Fig. \ref{BlowUp}. Besides the saturation of the gradient 
the graph illustrates the importance of the number of Fourier 
modes $K$ involved in the numerical simulation. The jumps 
in the graph correspond to the instances when we doubled $K$ 
in order to match the continuously increasing size of the flame. 
The graph is getting smoother if $K$ is adjusted more gradually 
and is kept much larger than the ratio of the critical wavelength 
$\lambda_{cr}=4\pi/\gamma$ to the circumferential length of 
the flame. The graph on the right of Fig. \ref{BlowUp} 
depicts the energy of the highest order Fourier harmonic 
involved in the simulation.

According to the data obtained in numerical simulations we 
may summarize the mechanism of flame front expansion in the 
Fourier space as follows. The energy of relatively long wave harmonics 
$\lambda>\lambda_{cr}$ permanently grows according to the Darrieus-Landau 
instability. This gained energy is transferred towards 
shorter wavelengths via the nonlinear effects and dissipates 
through the modes of short enough wavelengths $\lambda<\lambda_{cr}$. 
As flame expands the number of short enough angular modes reduces 
and the overall dissipation rate may become insufficient to 
counterbalance the generation of energy due to the Darrieus-Landau 
instability. This results in an accumulation of energy in the short 
wavelength spectrum, see Fig. \ref{BlowUp} (right), and leads 
to the blow-up of the numerical solution if the number $K$ of used 
harmonics is not increased 
in time. This effect explains a slight elevation of the short 
wave tail of the spectral energy distribution on the right of 
Fig. \ref{SpecDens} in comparison to the graph on the left. The 
latter one corresponds to the time moment soon after $K$ was doubled, 
though the former one is just approaching the moment when $K$ 
needs to be increased. 

In general, the process of stabilization of the expanding 
spherical front to a saturated state is very similar to 
the planar flame. The only distinctive difference is that 
the transitional period for the expanding flame is much 
longer. Using the realistic set of dimensional parameters 
from \cite{Filyand-Sivashinsky-Frankel94}, e.g. planar 
flame speed relative to the burnt gases $u_{b}=0.5~m/sec$ 
and thermal diffusivity 
$D_{th}=2.5\times 10^{-5}~m^{2}/sec$, one may interpret 
our findings in dimensional terms as follows. First cusps 
begin to appear on the flame surface for 
$\xavr{r}\ \approx 0.1~m$; flame acceleration with the 
rate $\xavr{r}\ \propto t^{3/2}$ starts for 
$\xavr{r}\ \approx 0.4~m$; acceleration rate begins to 
slow down for $\xavr{r}\ \approx 5~m$, and the 
acceleration ceases for $\xavr{r}\ \approx 20~m$. Most of 
experiments summarized in 
\cite{Gostintsev-Istratov-Shulenin88} and reported in 
more recent works 
\cite{Bradley-Hicks-Lawes-Sheppard-Woolley98,Rozenchan-Zhu-Law-Tse02} 
were carried out in enclosures and were affected by 
essential pressure rise and acoustics. Thus, there might 
be a considerable discrepancy between the characteristic 
flame radii just reported and those measured in realistic 
combustion experiments. Eventually, of course there is a 
chance that the stabilization of the expansion rate is 
just a consequence of the perturbative nature of the 
governing model used in this work.

\subsection{Effect of forcing}
A random point-wise set of perturbations uniformly distributed 
in time and in the Fourier space is a suitable model for both 
the computational round-off errors and a variety of perturbations 
of physical origins. In general, such a model would look like  
\bequ
f(x,t)=\sum\limits_{m=1}^{M(t)}a_{m}\cos(\xi_{m}\phi+\varphi_{m})
\delta(t-t_{m}),
\eequ{noise1}
where $a_{m}$, $t_{m}$, $\xi_{m}$, and $\varphi_{m}$ are 
non-correlated random sequences. It is assumed that 
$t_{1}\le t_{2}\le\cdots\le t_{m}\le\cdots\le t_{M(t)}\le t$,
$0\le\varphi_{m}\le 2\pi$, and $\xi_{m}\ge 0$, $m=1,2,\ldots,M(t)$.
However, in practice we use only two harmonics with 
$\xi=\xi_{*}=\floor\left[\gamma\!\xavr{r}\!\!/2\right]$ 
and $\xi=\xi_{*}+1$
weighted according to their closeness to the critical 
wavenumber $\xi_{cr}=\gamma\!\xavr{r}\!\!/2$: 
\bequ
f(x,t)\approx f_{0}[(1-\xi_{cr}+\xi_{*})\cos\xi_{*}\phi
+(\xi_{cr}-\xi_{*})\cos(\xi_{*}+1)\phi]
\sum\limits_{m=0}^{\infty}\delta(t-t_{m}).
\eequ{noise2}
These two harmonics approximate the critical planar flame 
harmonic of the wavelength $\lambda_{cr}=4\pi/\gamma$, 
which was shown to contribute the most in the fastest growing 
perturbations in the Sivashinsky-type models of flame 
dynamics, see \cite{Karlin05}.

The sign of the noise amplitude $f_{0}$ in \eq{noise2} 
was either plus or 
minus for every $m$ with the probability $1/2$. 
The delta function $\delta(t-t_{m})$ was approximated by 
$(\pi\tau)^{-1/2}e^{-(t-t_{m})^{2}/\tau}$ with a 
small enough value of $\tau$.
The impulse-like noise \eq{noise2} is used here for the 
sake of simplicity. Some arguments towards its validity were 
suggested in \cite{Joulin88}. Validation of models like \eq{noise1}, 
\eq{noise2} and of the present numerical approach was carried 
out in \cite{Karlin05}. More sophisticated and 
physically realistic models of temporal noise characteristics 
can be used with \eq{SivCurv1b} as well.

\begin{figure}[ht]
\begin{centering}\hfill
\begin{tabular}{p{51mm}p{0mm}p{71mm}}
\includegraphics[height=35mm,width=51mm]{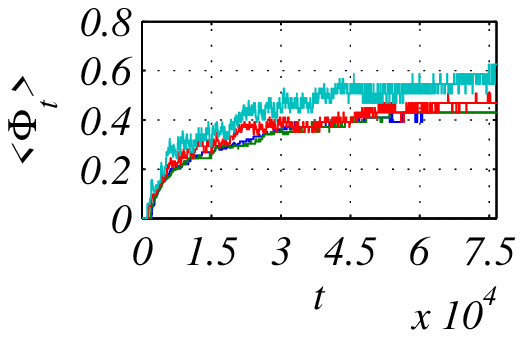} 
& &
\includegraphics[height=35mm,width=71mm]{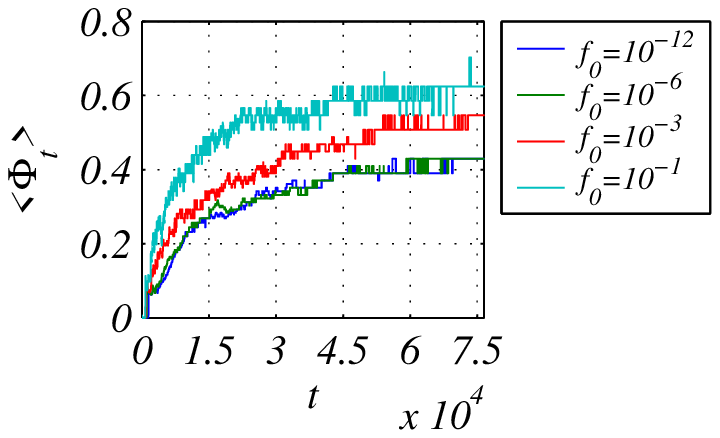}\\
\end{tabular}\hfill
\end{centering}
\caption{Effect of forcing amplitude and frequency on the 
averaged flame expansion speed for 
$\xavr{t_{m+1}-t_{m}}=50$ (left) and $\xavr{t_{m+1}-t_{m}}=10$ 
(right).}
\label{ForceEffect}
\end{figure}

The importance of the effect of the amplitude of noise and 
frequency of appearance 
of the impulse-like perturbations on the flame speed is 
illustrated in Fig. \ref{ForceEffect}. 

\begin{figure}[ht]
\begin{centering}\hfill
\begin{tabular}{p{51mm}p{0mm}p{71mm}}
\includegraphics[height=35mm,width=51mm]{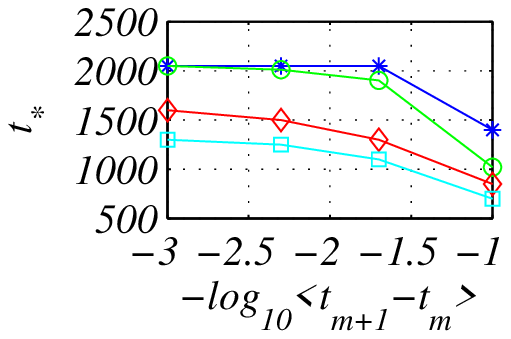} 
& &
\includegraphics[height=35mm,width=71mm]{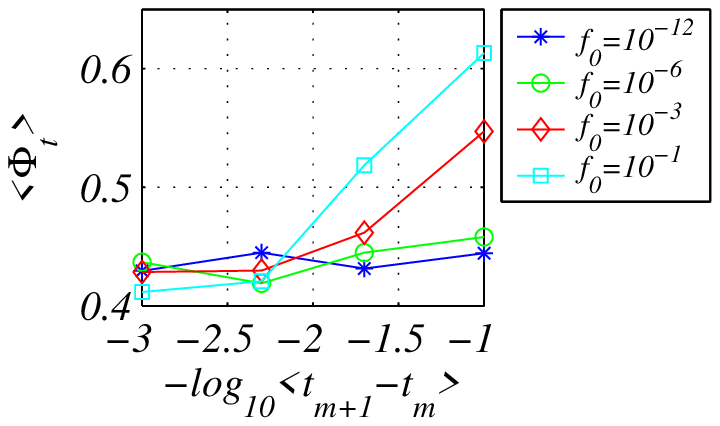}\\
\end{tabular}\hfill
\end{centering}
\caption{Effect of forcing amplitude and frequency on the 
beginning of acceleration (left) and stabilized flame expansion 
speed (right).}
\label{Noise}
\end{figure}

More details of the effect of noise are presented in 
Fig. \ref{Noise}. On the left we plotted graphs of the time 
instance at which the flame begins to accelerate versus frequency 
of appearance of the impulse-like perturbations for a variety 
of the perturbation amplitude $f_{0}$. Similar graphs of 
the nearly saturated flame expansion speed are given on the 
right.

\subsection{Three-dimensional numerical experiments}
In order to integrate \eq{GovEq3}-\eq{GovEq2b} we used an 
algorithm which is very similar to that one developed in 
\cite{DAngelo-Joulin-Boury00}. In addition, we parallelized 
the algorithm and used a few techniques to improve its stability 
and accuracy, see \cite{Mai-Karlin05}.

The basis functions in the spherical harmonics transformation 
(SHT) are the orthonormal eigenfunctions of the Laplace operator 
in spherical coordinates. However, unlike the Discrete 
Fourier Transformation (DFT), the discrete SHT of a data set is 
its approximation rather than an equivalent representation. 
Namely, it links a full matrix with $N^{2}$ elements representing 
a function on a regular spherical mesh of $N^{2}$ nodes in 
physical space and a triangular matrix of $N^{2}/2$ \shc. 
Thus, every back/forward cycle results in loss of information. 
This loss is similar to the effect of a short wave filter. In 
our approach it is associated with the nonlinear term only.

Coordinate singularities at the sphere poles result in 
accumulation of approximation errors near the poles and 
weaken numerical stability. In order to prevent accumulation 
of approximation errors near the poles, we rotate the 
coordinate system around an axis in the equatorial plane by 
an angle $\omega$ from time to time. The coordinate 
transformation formulas for such an axis passing through the 
points $\phi=\pm\pi/2$ are 
\[
\cos\theta^{\prime}=\cos\theta\cos\omega+\sin\theta\sin\omega\cos\phi
\]                                                                                   
\[
\cot\phi^{\prime}=\cot\phi\cos\omega-\csc\phi\sin\omega\cot\theta
\]                                                                                   
Their structure precludes use of the addition theorem for 
spherical polynomials to transform the coefficients 
$\widetilde{\Phi}_{n,m}$ to the new coordinate system without 
the global SHT back to the physical space. On the other hand, 
the back/forward SHT entails the application of a short wave filter 
to the whole solution $\widetilde{\Phi}_{n,m}$, rather than 
just to the nonlinear term $\widetilde{N}_{n,m}$. Hence, these 
rotations should not be done too frequently. Also, it is useful 
to combine them with rotations around the axis passing through 
the poles. Implementation of the latter ones is trivial.

Message Passing Interface (MPI) parallelization paradigm was 
implemented to allow the computational work to be distributed 
to a number of processors. However, the distributed data needs 
to be exchanged between these processors, which creates the biggest 
problem in modelling dynamics of large radius flames.
Unlike the multidimensional DFT, discrete SHT does not possess 
a structure of a tensor product of one-dimensional 
transformations. Therefore, the data transmission required 
by the truly distributed discrete SHT is much more 
sophisticated and intensive than just back/forward row/column 
transposition of the global solution array required by the DFT, 
see e.g. \cite{Karlin-Maz'ya-Schmidt03}. This results in a very 
fast growth of the communication overheads when the number of 
processors increases.

\begin{figure}[ht]
\begin{centering}\hfill
\begin{tabular}{p{60mm}p{1mm}p{60mm}}
\includegraphics[height=58mm,width=60mm]{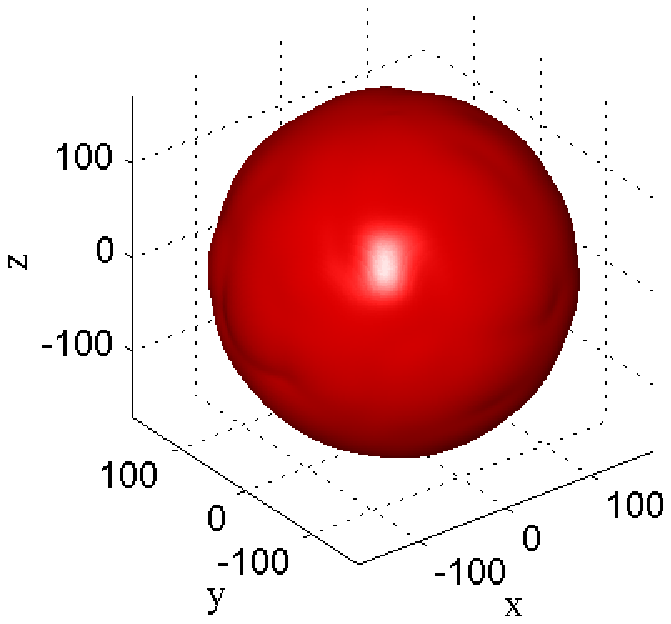}
& \phantom{a} &
\includegraphics[height=58mm,width=60mm]{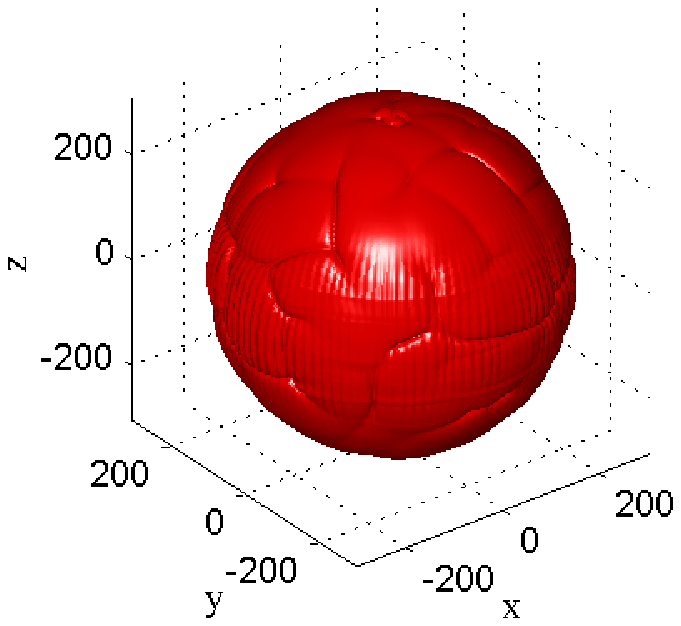}\\ 
\end{tabular}\hfill
\end{centering}
\vspace{-5pt}
\caption{Evolution of a spherical flame; $t\approx 40$ (left) 
and $t\approx 72$ (right). Here $\gamma=0.764$.}
\label{sphere3d}
\end{figure}

Spherical computational algorithms based on parallel 
Legendre-Fourier transformations are much less efficient than 
the standard planar Fourier methods. However, expanding 
spherical flames can be successfully simulated by the parallel 
SFT method. An example of the evolution of a random 
three-dimensional perturbation of a spherical flame is 
illustrated in Fig. \ref{sphere3d}. 

\subsection{Dynamics of curved flame segments}

Our numerical experiments showed that long time simulations of 
the three-dimensional expanding flames using discrete SHT might 
be possible on a parallel computer with large enough physically shared 
memory. However, there is an alternative approach based on 
simulations of 
the three-dimensional flame segments and following extension to 
the whole surface by periodicity. In order to validate this idea 
we applied it to the two-dimensional flames first. 

\begin{figure}[ht]
\begin{centering}\hfill
\begin{tabular}{p{51mm}p{0mm}p{71mm}}
\includegraphics[height=35mm,width=51mm]{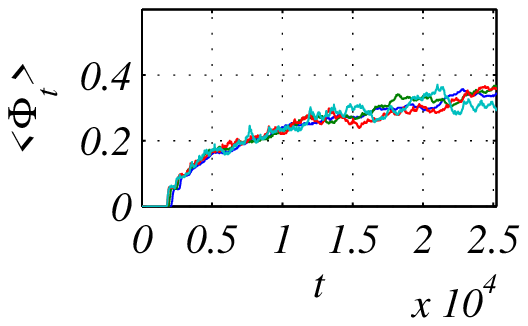} 
& &
\includegraphics[height=35mm,width=71mm]{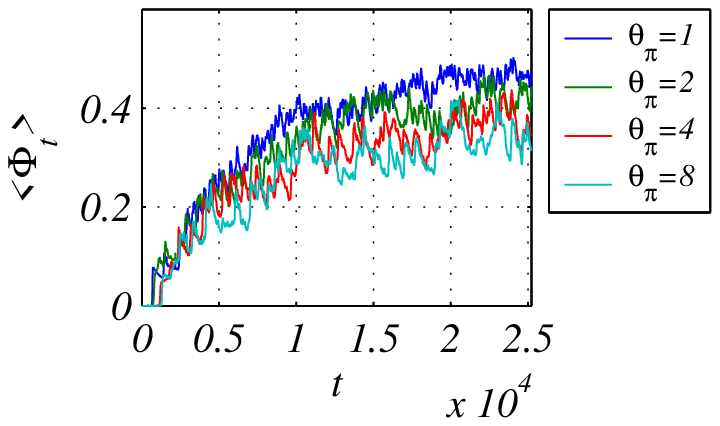}\\
\end{tabular}\hfill
\end{centering}
\caption{Averaged flame segment expansion speed for 
$f_{0}=0$ (left) and $f_{0}=0.1$, $\xavr{t_{m+1}-t_{m}}=10$ 
(right).}
\label{ForceEffectS}
\end{figure}

Results of two-dimensional numerical simulations of the dynamics 
of sectors $\theta\in[0,2\pi/\theta_{\pi}]$ of a spherical flame 
are illustrated in Fig. \ref{ForceEffectS}. One may see that the 
narrowing of the sector does not affect the flame expansion 
rate in absence of explicit forcing and that a correlation 
between this rate and the size of the segment becomes apparent 
as forcing strengthens. Similar observations were obtained 
for other parameters discussed earlier in this Section.

Our calculations also show that there is a critical value 
of $\theta_{\pi}$ above which sectorial simulations no longer 
represent the whole spherical flame. For example, we would not 
present the results of our simulations with $\theta_{\pi}=16$ as 
an approximation for the whole spherical flame. However, based 
on our two-dimensional results, it looks like simulations 
of the three-dimensional spherical flames using the Fourier, 
rather than the Fourier-Legendre, spectral model of the 
Sivashinsky type in the sector $0\le\theta,\phi\le\pi$, or 
even in $0\le\theta,\phi\le\pi/2$, are safe.

\section{Conclusions}
Long time interval simulations of a simplified model of 
the expanding spherical 
flames indicated that their expansion rate slows down as 
the flame size grows. The saturation of the planar flame 
propagation speed as their size grows was established too. 
Hence, a hypothesis of stabilization of the spherical flame 
expansion rate over a finite time interval is proposed. 

Further similarities with the propagating planar flames 
achieved by the expanding spherical flames on large time 
intervals were established in the studies of the effect of 
forcing, revealing a clear correlation between the strength 
of the forcing and the flame expansion rate. This supports 
the idea that the acceleration of both planar and expanding 
flames results from explicit and/or implicit forcing, which 
is always present both in experiments and calculations at 
least as noise of various physical origins.

In spite of many benefits of the simplified flame dynamics 
models, they have not been constructed to cope with 
significant perturbations of spherical flames. This casts 
a reasonable doubt in the possibility of extending the effects  
observed for the simplified model to realistic flames. 
In order to verify the hypothesis, numerical simulations of 
a more sophisticated model are required. In particular, 
model \cite{Frankel90} is valid for flames of any 
geometry if thermal gas expansion due to combustion does 
not generate 
significant vorticity, i.e. for $\gamma\ll 1$. However, 
even in the coordinate form \eq{int5}-\eq{int5c} the 
governing equation of model \cite{Frankel90} is extremely 
difficult to solve numerically, because, in contrast to 
\eq{SivCurv1c}, its nonlocal term has no convolution 
structure.

Using parallelized spherical harmonics transformation, the 
evolution of a three-dimensional expanding spherical flame 
has been successfully simulated to a stage when wrinkles 
appear and form a well developed cellular structure. 
However, computational problems associated with the 
spherical harmonics transformation make it difficult to 
extend these calculations on time intervals which would match 
those attained in our two-dimensional simulations. On the other 
hand, it was also noticed that the simulations of the 
three-dimensional spherical flames using the Fourier, 
rather than the Fourier-Legendre, spectral model of the 
Sivashinsky type in the sector $0\le\theta,\phi\le\pi$, or 
even in $0\le\theta,\phi\le\pi/2$, are reasonable.

\subsection*{Acknowledgements}
This work was supported by the EPSRC (Grant GR/R66692, UK), 
the US-Israel Binational Science Foundation (Grant 200 2008), 
the German-Israel Foundation (Grant G-695-15.10/2001), and 
the Israel Science Foundation (Grants 67-01 and 278-03). One 
of the authors is grateful to the Royal Society, The 
Academy Study Group on Israel and the Middle East, and to 
the Tel Aviv University for their support when working on 
this paper.

\end{document}